\documentstyle[12pt,epsfig]{article}                                                    

\newlength{\dinwidth}                                                    
\newlength{\dinmargin}                                                    
\def\lapproxeq{\lower .7ex\hbox{$\;\stackrel{\textstyle                                                    
<}{\sim}\;$}}                                                    
\def\gapproxeq{\lower .7ex\hbox{$\;\stackrel{\textstyle                                                    
>}{\sim}\;$}}                                                    
\def\be{\begin{equation}}                                                    
\def\ee{\end{equation}}                                                    
\def\bea{\begin{eqnarray}}                                                    
\def\eea{\end{eqnarray}}                                                    
\def\GeV{\rm GeV}

\begin{document}                                                    
\titlepage                                                    
\begin{flushright}                                                    
IPPP/08/52   \\
DCPT/08/104 \\                                                    
\today \\                                                    
\end{flushright}                                                    
                                                    
\vspace*{2cm}                                                    
                                                    
\begin{center}                                                    
{\Large \bf Diffractive dissociation re-visited for predictions at the LHC}                                                    
                                                    
\vspace*{1cm}                                                    
E.G.S. Luna$^{a}$, V.A. Khoze$^{a,b}$, A.D. Martin$^a$ and M.G. Ryskin$^{a,b}$ \\                                                    
                                                   
\vspace*{0.5cm}                                                    
$^a$ Institute for Particle Physics Phenomenology, University of Durham, Durham, DH1 3LE \\                                                   
$^b$ Petersburg Nuclear Physics Institute, Gatchina, St.~Petersburg, 188300, Russia            
\end{center}                                                    
                                                    
\vspace*{2cm}                                                    
                                                    
\begin{abstract}                                                    
We describe the formalism, and present the results, for a triple-Regge analysis of the available $pp$ and $p\bar{p}$ high-energy data which explicitly accounts for absorptive corrections. In particular, we allow for the gap survival probability, $S^2$, in single proton diffractive dissociation. Since for $pp$ scattering the value of $S^2$ is rather small, the triple-Pomeron vertex obtained in this analysis is larger than that obtained in the old analyses where the suppression caused by the absorptive corrections was implicitly included in an {\it effective} vertex. We show that the {\it bare} triple-Pomeron coupling that we extract from the $pp$ and $p\bar{p}$ data is consistent with that obtained in a description of the $\gamma p \to J/\psi+Y$ HERA data. The analyses of the data prefer a zero slope, corresponding to the small size of the bare vertex, giving the hope of a smooth matching to the perturbative QCD treatment of the triple-Pomeron coupling.  
\end{abstract}    

\section{Motivation}

The total and elastic proton-proton cross sections are usually described in terms of an eikonal model \cite{block}. The advantage of using an eikonal framework is that it automatically satisfies $s$-channel unitarity, which follows once we include the elastic rescattering of the interacting particles. Furthermore, the Good-Walker formalism \cite{GW} allows the possibility of excited proton states occurring in the intermediate state. In this way we include low-mass diffractive dissociation. It was demonstrated in \cite{newsoft} that a two-channel eikonal is sufficient to capture the main features of this dissociation (provided, of course, we are not interested in the production of a particular resonant state), see also \cite{GLMnew}.

On the other hand, high-mass ($M$) diffractive dissociation is described in terms of the triple-Regge formalism, where the differential cross section, $d\sigma/dtdM^2$, is driven by the triple-Regge couplings $g_{R_1R_2R_3}$. The values of these couplings were determined from the data available in triple-Regge domain in the 1970s; see, for example \cite{KKPI,FF,Kaid}. However, these early analyses did not allow for absorptive corrections, which are sizeable in hadron-hadron collisions. Therefore, the old triple-Regge couplings must be regarded, not as the bare vertices, but as effective couplings embodying absorptive effects. It was pointed out in Ref.~\cite{capella} that the original {\it bare} triple-Pomeron vertex may be about three times larger than the {\it effective} coupling which is obtained with the neglect of absorptive corrections. However, absorptive effects are very sensitive to the $t$-slope of the triple-Reggeon vertices and in the 1970s it was ``impossible to determine the exact values of the {\it bare} triple-Pomeron coupling'' \cite{capella}; the accuracy and energy range covered by the data were insufficient. So the results of the triple-Regge analyses were presented in the form of {\it effective} couplings. This leads to a problem, since the absorptive effects are not constant factors, but depend on the energy and the type of diffractive process. Since the inelastic cross section expected at the LHC is more than twice as large as that observed at fixed target and CERN-ISR energies, we cannot use the results of the old triple-Regge analyses to predict the diffractive effects at the LHC\footnote{Already, in order to describe data up to the Tevatron energy, Goulianos and Montanha \cite{GM} found it necessary to restore unitarity by renormalising the ``Pomeron flux'' \cite{renorm}. This mimics to some extent the absorptive effects.}. It is therefore necessary to perform a new triple-Regge analysis that includes the absorptive effects explicitly. Here we present the first analysis of the available diffractive (and elastic) data in the CERN-ISR -- Tevatron region in which the absorptive corrections are systematically included. The absorptive corrections are calculated in terms of a two-channel eikonal model fitted to describe the total and elastic differential cross section data for $pp$ and $p\bar{p}$ scattering. 

At the moment, the energy behaviour of the scattering amplitude
may be consistently described by two different scenarios for the asymptotic regime \cite{bh75}.
One is called the {\it weak coupling} of the Pomerons. In this case,
at very high energy, the cross sections tend to the
universal constant value
\be
\sigma_{\rm tot} \to {\rm constant}~~~~~~{\rm as}~~~~s \to \infty.
\ee
In order not to violate unitarity, the triple-Pomeron coupling must vanish with vanishing transverse momentum transferred through the Pomeron \cite{GribMYF}
\be
g_{3P}~\propto~ q^2_t~~~~~~{\rm as}~~~~q_t \to 0.
\label{eq:weak}
\ee
Another possibility is called the {\it strong coupling} scenario \cite{GribM}. Here, at a
very high energies, the cross sections grows as 
\be
\sigma_{\rm tot}\propto {({\rm ln}~s)}^\eta ~~~~~{\rm with}~~~~~ 0<\eta \leq 2,
\ee 
and the bare vertex 
\be
g_{3P}|_{q_t \to 0}~~\to~~{\rm constant}.
\label{eq:strong}
\ee

The present data are usually described within the
Froissart-like limit of the second scenario (with $\eta=2$).
However to reach asymptotics we need {\bf very} high energy --
the energy at which the slope of the elastic amplitude,
$B=B_0+\alpha'_P{\rm ln}(s)$ is dominated by the second term, that is
when $\alpha'_P{\rm ln(s)} \gg B_0$. This is far beyond the energies
available at present. Another possibility, to distinguish between the {\it
weak} and {\it strong} approaches, is to study the $q_t$ dependence of the
bare triple-Pomeron vertex. Thus, it is important to extract the {\it
bare} vertex before its behaviour is affected by absorptive
corrections.

This paper is devoted to a detailed analysis of ``soft'' scattering data using, in turn, the {\it strong} and {\it weak} triple-Pomeron coupling behaviours. In Section 2 we recall
the Good-Walker formalism for the two-channel eikonal model,
 in Section 3 we briefly describe the parametrisation used to
calculate the absorptive corrections for the triple-Reggeon cross
sections and demonstrate that this parametrisation is consistent with
the available data on the differential elastic ($d\sigma_{\rm el}/dt$) and
total cross sections.  Next, in Section 4, we give the formulae for
the inclusive cross section, $M^2d\sigma/dtdM^2$, in the triple-Regge
region which incorporates the screening corrections caused by the
two-channel eikonal. The results of our triple-Regge analyses of
$pp\to pX$ and $\bar{p}p \to \bar{p}X$ data are presented in Section 5. We perform fits using both the strong and weak triple-Pomeron coupling. We find that the data favour the former ansatz. In Section 6 we discuss the description of the data for
inelastic diffractive $J/\psi$ photoproduction, $\gamma p\to J/\psi+Y$, obtained at HERA.
We find that the $J/\psi$ HERA data again favour the strong triple-Pomeron coupling scenario.
In Section 7 we present our conclusions.

\section{R\'{e}sum\'{e} of the eikonal formalism}

\subsection{Single-channel eikonal model}

First, we briefly recall the relevant features of the single-channel eikonal model. That is we focus on elastic
unitarity. Then ``disc $T$'' is simply the discontinuity of the $pp$ scattering amplitude $T$
across the two-particle $s$-channel cut. At high energies we have a sizeable inelastic component. The
$s$-channel unitarity relation is diagonal in the impact
parameter, $b$, basis, and may be written
\be 2 {\rm Im}\,T_{\rm el}(s,b) = |T_{\rm el}(s,b)|^2 + G_{\rm
inel}(s,b) \label{eq:a1} \ee
with
\bea \sigma_{\rm tot} & = & 2\int d^2b\, {\rm Im}\,T_{\rm el}(s,b) \label{eq:ot} \\
\sigma_{\rm el} & = & \int d^2b\,|T_{\rm el}(s,b)|^2 \\
\sigma_{\rm inel} & = & \int d^2b\,\left[2{\rm Im}\,T_{\rm
el}(s,b) - |T_{\rm el}(s,b)|^2\right]. \eea
These equations are satisfied by
\bea {\rm Im}T_{\rm el}(s,b) & = & 1-{\rm e}^{-\Omega/2} \label{eq:elastamp}\\
\sigma_{\rm el}(s,b) & = & (1-{\rm e}^{-\Omega/2})^2, \label{eq:el}\\
\sigma_{\rm inel}(s,b) & = & 1- {\rm e}^{-\Omega}, \label{eq:inel} \eea
where $\Omega(s,b)\geq 0$ is called the
opacity (optical density) or eikonal\footnote{Sometimes $\Omega/2$
is called the eikonal; for simplicity we omit the real part of $T_{\rm el}$. At high energies, the ratio ${\rm ReT}_{\rm el}/{\rm Im}T_{\rm el}$ is small, and can be evaluated via a dispersion relation.}. From (\ref{eq:inel}), we see that $ \exp(-\Omega(s,b))$ is the probability that
no inelastic scattering occurs.

  The opacity corresponding to an individual Regge pole $i=R$ or $P$ takes the form     
\be    
\label{eq:append9}    
\Omega_i (s, b) ~=~\frac{1}{\pi}\int d^2 q_t~A_i(s,q_t)e^{i\vec{q}_t\cdot\vec{b}} \; = \; \frac{\beta_i(0)^2 (s/s_0)^{\alpha_i(0)-1}}{B_i} \: e^{-b^2/4B_i},    
\ee
where $q_t^2=-t$, and where the contribution to the amplitude from Reggeon $i$ exchange is\footnote{Note
that, unlike
\cite{newsoft,oldsoft,KKMR}, here we use the normalisation
$\sigma^{\rm tot}_i=4\pi{\rm Im}\eta_i ~\beta_i(0)^2(s/s_0)^{\alpha_i(0)-1}$,
rather than
$\sigma^{\rm tot}_i={\rm Im}\eta_i ~\beta_i(0)^2(s/s_0)^{\alpha_i(0)-1}$.
That is, as compared to our previous papers, here all the
coupling constants ($\beta_i$ and $g_{iij}$) are decreased by a
factor $\sqrt{4\pi}$.}
\be
A_i(s,q_t)~=~\eta_i \beta_i(0)^2 e^{r_i t} (s/s_0)^{\alpha_i(t)-1},
\label{eq:regge}
\ee    
where the $t$ behaviour of the Reggeon-proton vertex is taken to have the exponential form  
\be
\beta_i(t)~=~\beta_i(0) \exp (r_i t/2).
\label{eq:betaR}
\ee
Thus, for a linear trajectory
\be
\alpha_i=\alpha_i(0)+\alpha_i^\prime t,
\label{eq:lin}
\ee    
the $t$-slope of the amplitude $A_i$ is given by
\be    
\label{eq:append10}    
B_i \; = \; r_i \: + \: \alpha_i^\prime \: \ln (s/s_0).    
\ee
For Regge exchange with signature $\pm 1$, the factor $\eta_i$ in eq. (\ref{eq:regge}) is 
\be
\eta_i~=~\frac{-(1\pm e^{-i\pi\alpha_i})}{{\rm sin}\pi \alpha_i}.
\label{eq:sigfac}
\ee

Although we take exponential forms for the exchange of the secondary Reggeons, we choose, as in Ref.~\cite{oldsoft}, a power-like form for the proton-Pomeron vertex
\be
\beta_P(t)~=~\frac{\beta_P(0)}{(1-t/a_1)(1-t/a_2)}.
\label{eq:betaP}
\ee
Also, to improve the description of the low $|t|$ region, we follow Anselm and Gribov \cite{AG} and include pion-loop insertions\footnote{Note that, in (\ref{eq:a10}) below, we have corrected the misprint which occurs in the published version of \cite{AG}.} in Pomeron exchange which results in a non-linear form of the Pomeron trajectory    
\be    
\label{eq:a9}    
\alpha_P (t) \; = \; \alpha_P (0) \: + \: \alpha_P ^\prime t \: - \: \frac{\beta_\pi^2 m_\pi^2}{32     
\pi^3} \: h \left ( \frac{4 m_\pi^2}{| t |} \right ),    
\ee    
where    
\be    
\label{eq:a10}    
h (\tau) \; = \; \frac{4}{\tau} \: F_\pi^2 (t) \: \left [ 2 \tau \: - \: (1 + \tau)^{3/2} \: \ln \left (     
\frac{\sqrt{1 + \tau} + 1}{\sqrt{1 + \tau} - 1} \right ) \: + \: \ln \frac{m^2}{m_\pi^2} \right ],    
\ee    
with $\tau = 4m_\pi^2/|t|$ and $m = 1$~GeV.  The coefficient $\beta_\pi^2$ specifies the     
$\pi\pi$ total cross     
section,     
and $F_\pi (t)$ is the form factor of the pion-Pomeron vertex.  The coefficient $\beta_\pi^2     
m_\pi^2/32 \pi^3$ in (\ref{eq:a9}) is small\footnote{We use the additive quark model relation $\beta_{\pi}(\equiv g_{\pi\pi P})=\frac{2}{3}\beta_P(0)$.}, but due to the tiny scale $m_\pi$ the $t$     
dependence of $h (\tau)$ is steep and non-linear.  It has an important effect on the local slope, $d (\ln d\sigma_{\rm el}/dt)/dt$, of the elastic cross section. For the results that we obtain below for the Pomeron trajectory, $\alpha_P (t)$, it is     
important to note that expression (\ref{eq:a10}) for $h (\tau)$ has been renormalised     
\cite{AG}, such that    
\be    
\label{eq:b10}    
h (\tau) \; = \; h_\pi (\tau) \: - \: h_\pi (0)    
\ee    
where $h_\pi (\tau)$ denotes the full pion-loop contribution. Further discussion of the pion-loop contribution can be found in Ref. \cite{oldsoft}.

Finally, the total opacity is given by the sum of all the allowed Regge exchanges
\be
\Omega~=~\sum_{i=P,R}\Omega_i.
\ee    

In the present simplified analysis we neglect the real part of the amplitude, since our goal is to investigate the role of the absorptive effect which is caused essentially by the imaginary part of the amplitude, that is by the real part of the opacity. Therefore, the values of the couplings $\beta_i(0)$, that we quote in Table 1 below, are actually ${\rm Im}(\eta_i\beta_i(0))$, where the $\eta_i$ are the signature factors of (\ref{eq:sigfac}).

\subsection{Inclusion of low-mass diffractive dissociation}
So much for elastic diffraction. Now we turn to inelastic
diffraction, which is a consequence of the {\em internal
structure} of hadrons. This is simplest to describe at high
energies, where the lifetime of the fluctuations of a fast hadron is
large, $\tau\sim E/m^2$, and during these time intervals the
corresponding Fock states can be considered as `frozen'. Each
hadronic constituent can undergo scattering and thus destroy the
coherence of the fluctuations. As a consequence, the outgoing
superposition of states will be different from the incident
particle, and will most likely contain multiparticle states, so we
will have {\em inelastic}, as well as elastic, diffraction.

To discuss inelastic diffraction, it is convenient to follow Good
and Walker~\cite{GW}, and to introduce states $\phi_k$ which
diagonalize the $T$ matrix. Such eigenstates only undergo elastic
scattering. Since there are no off-diagonal transitions,
\be \langle \phi_j|T|\phi_k\rangle = 0\qquad{\rm for}\ j\neq k, 
\ee
a state $k$ cannot diffractively dissociate into a state $j$. We
have noted that this is not, in general, true for hadronic states, which are not eigenstates of the $S$-matrix, that is of $T$. To account for the internal structure of the hadronic states, we have to enlarge the set of
intermediate states, from just the single elastic channel, and to
introduce a multichannel eikonal. We will consider such an example
below, but first let us express the cross section in terms of the
probability amplitudes $F_k$ of the hadronic process proceeding via the
various diffractive eigenstates\footnote{The exponent exp$(-\Omega_k)$ describes the probability that the diffractive eigenstate $\phi_k$ is not absorbed in the interaction. Later we will see that the rapidity gap survival factors, $S^2$, can be described in terms of such eikonal exponents.} $\phi_k$.

Let us denote the orthogonal matrix which diagonalizes ${\rm
Im}\,T$ by $a$, so that
\be \label{eq:b3} {\rm Im}\,T \; = \; aFa^T \quad\quad {\rm with}
\quad\quad \langle \phi_j |F| \phi_k \rangle \; = \; F_k \:
\delta_{jk}. \ee
Now consider the diffractive dissociation of an arbitrary incoming
state
\be \label{eq:b4} | j \rangle \; = \; \sum_k \: a_{jk} \: | \phi_k
\rangle. \ee
The elastic scattering amplitude for this state satisfies
\be \label{eq:b5} \langle j |{\rm Im}~T| j \rangle \; = \; \sum_k
\: |a_{jk}|^2 \: F_k \; = \; \langle F \rangle, \ee
where $F_k \equiv \langle \phi_k |F| \phi_k \rangle$ and where the
brackets of $\langle F \rangle$ mean that we take the average of
$F$ over the initial probability distribution of diffractive
eigenstates. After the diffractive scattering described by
$T_{fj}$, the final state $| f \rangle$ will, in general, be a
different superposition of eigenstates from that of $| j \rangle$,
which was shown in~(\ref{eq:b4}). At high energies we may neglect
the real parts of the diffractive amplitudes. Then, for cross
sections at a given impact parameter $b$, we have
\bea \label{eq:b6} \frac{d \sigma_{\rm tot}}{d^2 b} & = & 2 \:
{\rm Im} \langle j |T| j \rangle \; = \; 2 \: \sum_k
\: |a_{jk}|^2 \: F_k \; = \; 2 \langle F \rangle \nonumber \\
& & \nonumber \\
\frac{d \sigma_{\rm el}}{d^2 b} & = & \left | \langle j |T| j
\rangle \right |^2 \; = \; \left (
\sum_k \: |a_{jk}|^2 \: F_k \right )^2 \; = \; \langle F \rangle^2 \\
& & \nonumber \\
\frac{d \sigma_{\rm el \: + \: SD}}{d^2 b} & = & \sum_k \: \left |
\langle \phi_k |T| j \rangle \right |^2 \; = \; \sum_k \:
|a_{jk}|^2 \: F_k^2 \; = \; \langle F^2 \rangle. \nonumber \eea
It follows that the cross section for the single diffractive
dissociation of a proton,
\be \label{eq:b7} \frac{d \sigma_{\rm SD}}{d^2 b} \; = \; \langle
F^2 \rangle \: - \: \langle F \rangle^2, \ee
is given by the statistical dispersion in the absorption
probabilities of the diffractive eigenstates. Here the average is
taken over the components $k$ of the incoming proton which
dissociates. If the averages are taken over the components of both
of the incoming particles, then in (\ref{eq:b7}) we must introduce a second index on $F$, that is $F_{ik}$, and sum over $k$ and $i$. In this case the sum is the
cross section for single and double dissociation.

\section{Description of the elastic data}

In preparation for the triple-Regge analysis of the data for high-mass diffraction, we first perform a two-channel eikonal fit to all the high-energy $pp$ and $p\bar{p}$ total cross section data above $\sqrt{s}=10$ GeV and to the elastic differential
scattering cross section for $pp$ at $\sqrt{s}=31$, 53 and 62 GeV, and for  $p\bar{p}$ at $\sqrt{s}=31$, 53, 62, 546 and 1800 GeV.
We use total cross section data sets compiled by the Particle Data Group \cite{PDG} and differential cross section data sets
from the references \cite{ISR} (ISR), \cite{SpS} ($Sp\bar{p}S$) and \cite{Tevatron} (Tevatron), see the review in Ref.~\cite{DG}. The statistic and systematic
errors of all scattering quantities have been added in quadrature.

We introduce a parameter $\gamma$ to define the two diffractive eigenstates $k$ of the eikonal model, such that their couplings to the Pomeron are
\be
\beta_{P,k}(t)~=~(1\pm \gamma) \beta_P(t),
\ee
where the proton wave function $ |p\rangle =(|+\rangle +|-\rangle)/\sqrt{2} $, see Ref.~\cite{oldsoft}. The CERN-ISR measurements of the excitations into particular channels ($N\pi,$ $N\pi\pi,$ $\Lambda K$ etc.) with $M<2.5$ GeV \cite{reson} correspond to a cross section for low-mass diffraction of\footnote{Here, and in what follows, the value of $\sigma_{\rm SD}$ accounts for dissociation of both colliding particles.}
\be
\sigma_{\rm SD}^{{\rm low}M} \simeq 2~{\rm mb~~~~~~~~at}~\sqrt{s}=31~{\rm GeV}. 
\label{eq:2mb}
\ee
This value corresponds to $\gamma \simeq 0.55$, which we take in this analysis.

We describe the $pp$ and $p\bar{p}$ total and elastic cross section data in terms of Pomeron, and positive ($f_2, a_2$) and negative ($\omega, \rho$) signature Regge exchange. The {\it positive} signature secondary Reggeons, $f_2$ and $a_2$, are taken to lie on an exchange-degenerate linear trajectory with intercept $\alpha_+(0)$ and coupling\footnote{If we assume the additive-quark-model relation, $\beta(f_2)=3\beta(a_2)$, then the coupling $\beta(f_2)=\sqrt{9/10}~\beta_+$.} to the proton of parametric form
\be
\beta_+(t)~=~\beta_+(0)~{\rm exp}(r_+t/2),
\ee
see (\ref{eq:betaR}). Similarly, the exchange-degenerate {\it negative} signature pair ($\omega, \rho$) are described by parameters $\alpha_-(0),~\beta_-(0)$ and $r_-$. The slopes, $\alpha'_+,~\alpha'_-$, of the secondary Reggeon linear trajectories are fixed at 0.9 ${\rm GeV}^{-2}$. On the other hand, the Pomeron coupling and trajectory are taken to have parametric forms given by (\ref{eq:betaP}) and (\ref{eq:a9}), respectively. 

\begin{figure}
\begin{center}
\includegraphics[height=.47\textheight]{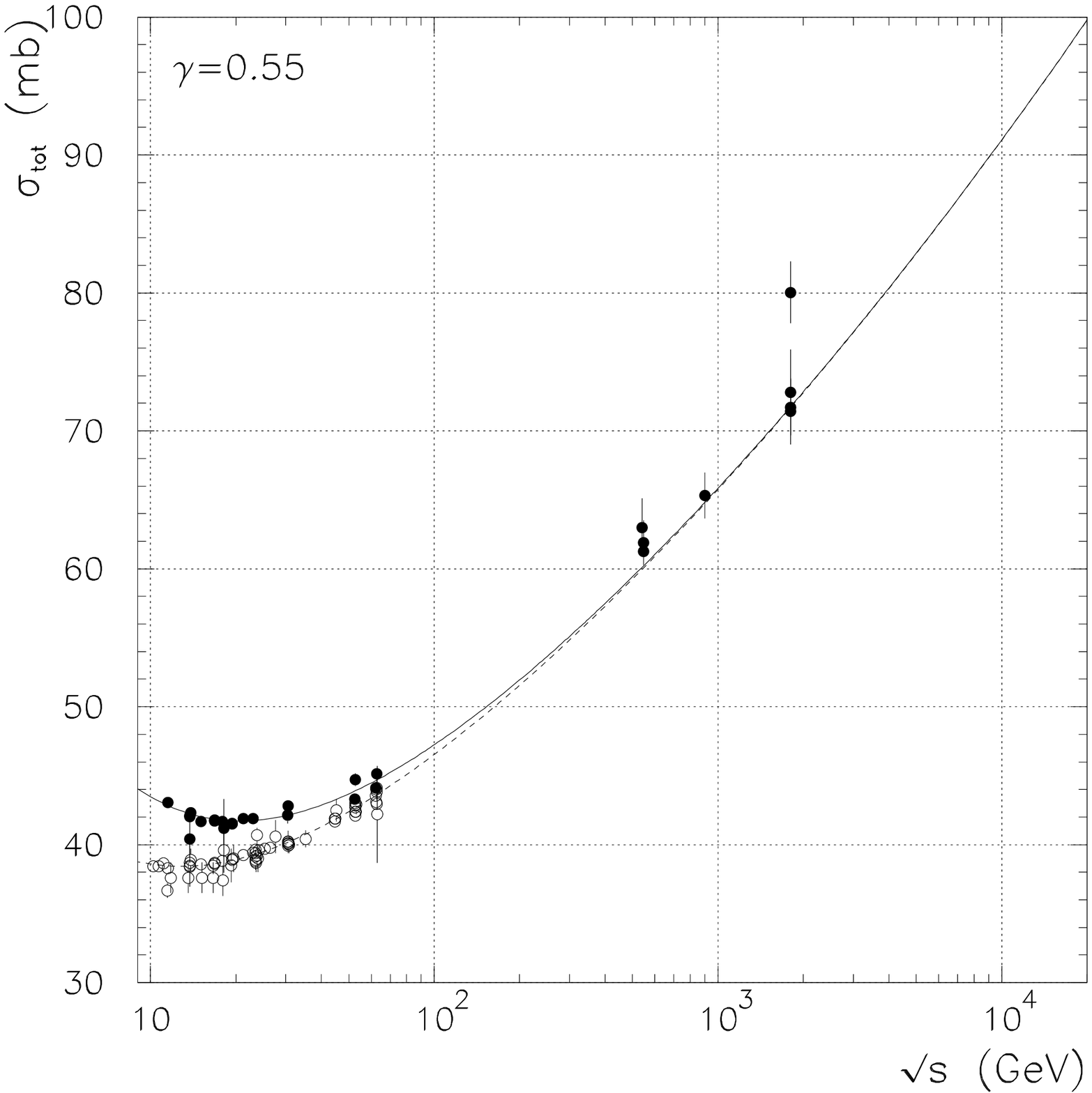}
\includegraphics[height=.47\textheight]{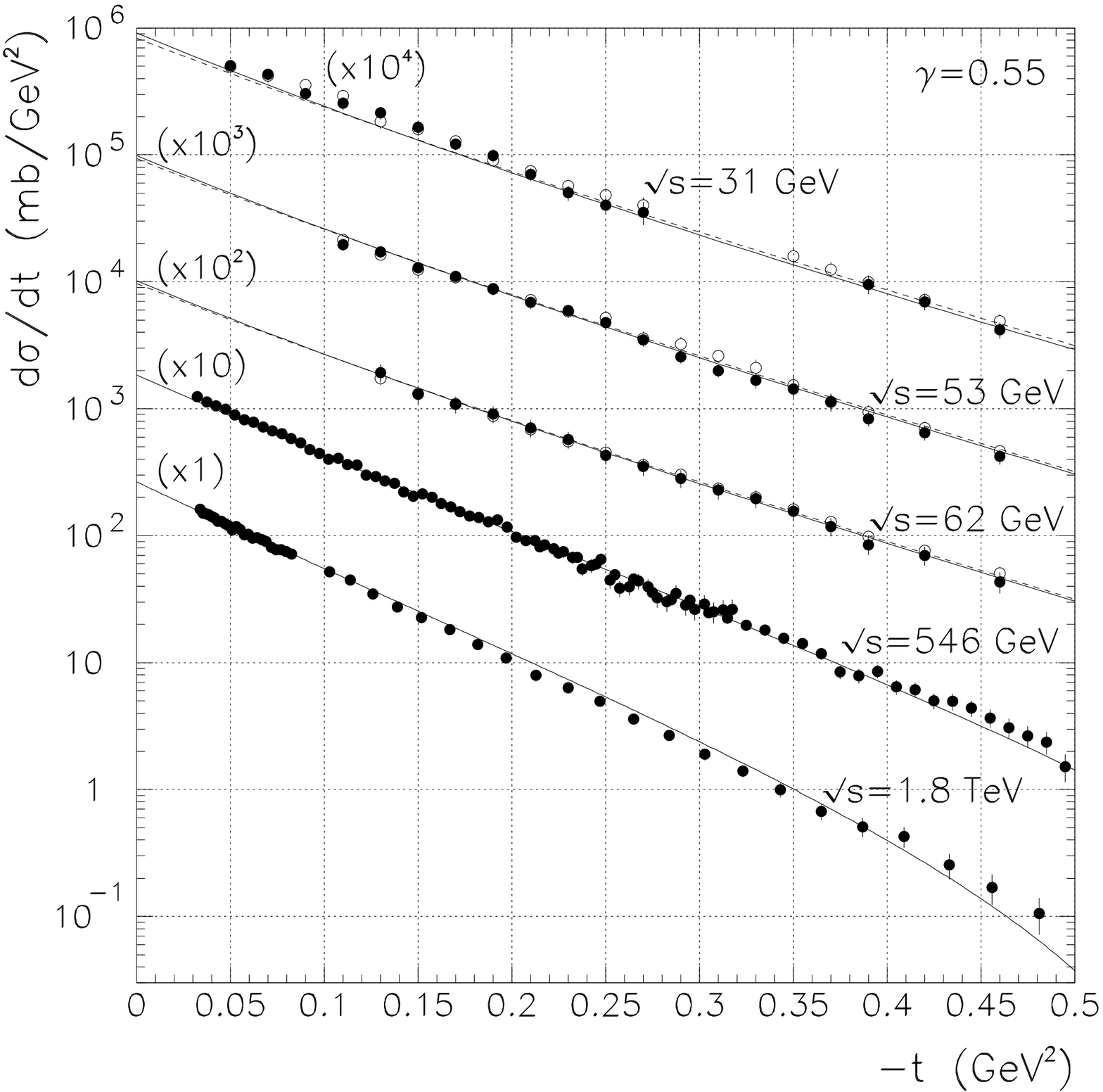}
\caption{\sf Two-channel eikonal model description of total and differential cross section data.}
\label{fig:elastic}
\end{center}
\end{figure}

The values of the Regge parameters determined by the global fit are listed in Table 1. The fit has $\chi^2/{\rm DoF}=259/(325-11)=0.83$. 
The description of $\sigma_{tot}$ and $d\sigma /dt$ is displayed in Fig.~\ref{fig:elastic}, where the solid (dashed) curves
correspond to $p\bar{p}$ ($pp$) scattering. The total cross
section at LHC is predicted\footnote{As expected, the inclusion of absorption reduces the value of the cross section predicted at the LHC energy. Indeed, if the high-mass diffractive absorptive effects were also included then the total cross section would be further reduced to about 90 mb \cite{newsoft,newnewsoft}, see also \cite{KGB,GLMnew}. The present model also predicts an elastic $pp$ differential cross
section at the LHC energy with a diffractive dip at $-t \simeq 0.4
\GeV^2$. However, for this prediction to be reliable we must await the
inclusion of the high-mass absorptive effects in the elastic analysis \cite{newnewsoft}. An earlier analysis \cite{newsoft}, which included high-mass diffraction
phenomenologically, showed that, at the LHC energy, the dip occurred at
larger $-t$ and that the elastic cross section was smooth in the region
shown in Fig.~\ref{fig:elastic}.} to be $\sigma_{\rm tot}= 94.8$ mb.
\begin{table}[htb]
\begin{center}
\begin{tabular}{|c|c|}\hline
$\beta_P(0)$  &  $2.26~\pm~0.02$   \\
$a_1$  &   $0.44 ~\pm ~0.01$  \\  
$a_2$  &   $16.3 ~\pm ~2.8$   \\
$\alpha_P(0)$  &   $1.121 ~\pm~0.001  $ \\
$\alpha'_P$  &   $0.033 ~\pm  ~0.002$ \\ \hline
$\beta_+(0)$  &   $5.9 ~\pm ~1.2 $ \\
$r_+$   &   $0.5 ~\pm ~1.2 $ \\
$\alpha_+(0)$  &   $0.54 ~\pm~0.04  $ \\  \hline
$\beta_-(0)$  &   $2.4 ~\pm~0.9  $ \\
$r_-$   &   $3.1 ~ \pm  ~6.0$ \\
$\alpha_-(0)$  &   $0.57 ~\pm ~0.09 $ \\  \hline
\end{tabular}
\end{center}
\caption{\sf The values of the Pomeron, and the positive and negative signature secondary Reggeon  parameters obtained in the fit to the total and elastic differential cross section $pp$ and $p\bar{p}$ data. GeV units are used; so, for example, the couplings $\beta(0)$ have units of ${\rm GeV}^{-1}$. The couplings $\beta_i(0)$ that we quote are actually ${\rm Im}(\eta_i\beta_i(0))$, where the signature factor $\eta_i$ is given by (\ref{eq:sigfac}), see the last paragraph of Section 2.1. The errors correspond to a 90$\%$ confidence level.}
\end{table}

\section{High-mass diffraction: triple-Regge formalism}
\begin{figure} 
\begin{center}
\includegraphics[height=3cm]{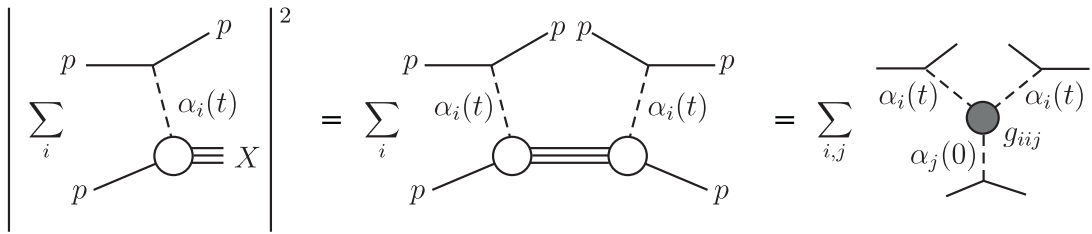}
\caption{\sf The triple-Regge description of high-mass diffractive dissociation, $pp \to pX$. In general, we may have vertices with three different Reggeons with coupling $g_{ii'j}$. However, here it is sufficient to consider only contributions with $i=i'$.}
\label{fig:3R}
\end{center}
\end{figure}
The multichannel eikonal is unable to account for diffraction into high mass states. These processes, $pp \to pX$ with large $M_X$, are usually described in terms of a triple-Regge formalism, where
\be
M_X^2~\simeq~(1-x_L)s,
\ee
where $x_L \equiv 1-\xi$ is the momentum fraction of the ingoing proton carried by the outgoing proton.
The approach is sketched in Fig. \ref{fig:3R}. Since we have large $M_X$ we no longer have $-t=q_t^2$. Rather, we must allow for non-vanishing $t_{\rm min}$
\be
-t~=~\frac{q_t^2}{x_L}+\frac{m_p^2(1-x_L)^2}{x_L}~=~\frac{q_t^2}{x_L}-t_{\rm min}.
\ee

To the best of our knowledge, the screening corrections have not been explicitly included in the triple-Regge formalism.  Therefore we present the formalism below, first using a single-channel eikonal and then generalising it to the two-channel case. Finally we treat the corrections to the $\pi\pi P$ diagram as a special case.

\subsection{Screening corrections in the triple-Regge formalism}
\begin{figure} 
\begin{center}
\includegraphics[height=4cm]{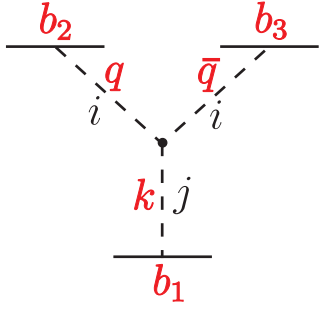}
\caption{\sf A schematic diagram showing the notation of the impact parameters arising in the calculation of the screening corrections to the $iij$ triple-Regge diagram. The conjugate momenta to $b_1,b_2,b_3$ are $k_t,q_t,\bar{q}_t$. If $k_t=0$, then $\bar{q}_t=q_t$.}
\label{fig:3Rb}
\end{center}
\end{figure}
If we first neglect the screening correction, then the $iij$ triple-Regge diagram of Fig. \ref{fig:3R} gives the contribution
\be
\frac{M^2 d\sigma}{dtdM^2}~=~\beta_j(0)\beta_i^2(t)g_{iij}(t)\left(\frac{s}{M^2}\right)^{2\alpha_i(t)-2}\left(\frac{M^2}{s_0}\right)^{\alpha_j(0)-1}.
\ee
We use a simple exponential parametrisation of the triple-Regge vertices
\be
g_{iij}(t)~=~g_{iij}(0)~{\rm exp}(b'_{iij}(q^2+\bar{q}^2-k^2_t)),
\ee
where the momenta are defined in Fig. \ref{fig:3Rb}, and where $q^2=t_{\rm min}-q^2_t/x_L$.

Screening effects are best included by working in impact parameter space and using suppression factors of the form exp($-\Omega(b)$). To determine the $q_t$ or $t$ dependence we take the Fourier transforms with respect to the impact parameters specified in Fig. \ref{fig:3Rb}. We obtain\footnote{Note that $e^{i\vec{k}_t \cdot \vec{b}_1}=1$ as $k_t=0$.}
\be
\frac{M^2 d\sigma}{dtdM^2}~=~A\int\frac{d^2b_2}{2\pi}e^{i\vec{q}_t \cdot \vec{b}_2} F_i(b_2)\int\frac{d^2b_3}{2\pi}e^{i\vec{q}_t \cdot \vec{b}_3} F_i(b_3)\int\frac{d^2b_1}{2\pi} F_j(b_1),
\label{eq:3Rb}
\ee
where
\be
F_i(b_2)~=~\frac{1}{2\pi \beta_i(q_t=0)}\int d^2q_t \beta_i(q_t)\left(\frac{s}{M^2}\right)^{-\alpha^\prime_i q^2_t} e^{b^\prime_{iij}q^2}e^{i\vec{q}_t \cdot \vec{b}_2},
\label{eq:Fi}
\ee
\be
F_j(b_1)~=~\frac{1}{2\pi \beta_j(k_t=0)}\int d^2k_t \beta_j(k_t)\left(\frac{M^2}{s_0}\right)^{-\alpha^\prime_j k^2_t} e^{-b^\prime_{iij}k^2_t},
\label{eq:Fj}
\ee
and where the $q_t$-independent factors are collected in $A$
\be
A~=~\beta_j(0)\beta_i^2(0)g_{iij}(0)\left(\frac{s}{M^2}\right)^{2\alpha_i(t_{\rm min})-2}\left(\frac{M^2}{s_0}\right)^{\alpha_j(0)-1}.
\label{eq:A}
\ee
These equations assume that we have a strong triple-Pomeron scenario, see (\ref{eq:strong}). For the weak triple-Pomeron coupling ansatz, (\ref{eq:weak}), we must include a factor $q_t$ in the integrand of the expression (\ref{eq:Fi}) for $F_i(b_2)$ when $i,j=P$, and also $\bar{q}_t~(=q_t)$ in the analogous formula for $F_i(b_3)$.

To include the screening corrections, for a single-channel eikonal, we must include in the integrands on the right-hand side of (\ref{eq:3Rb}) the factors
\be
{\rm exp}(-\Omega(\vec{b}_2-\vec{b}_1)/2)~{\rm exp}(-\Omega(\vec{b}_3-\vec{b}_1)/2)~\equiv~S(\vec{b}_2-\vec{b}_1)~S(\vec{b}_3-\vec{b}_1).
\ee
That is, we need to compute
\be
\left. \frac{M^2 d\sigma}{dtdM^2}\right|_{iij}~=~A\int\frac{d^2b_1}{2\pi} F_j(b_1) |I_d(b_1)|^2,
\label{eq:3Rbscreen}
\ee
where $I_d$ is given by
\be
I_d(b_1)~\equiv~\int\frac{d^2b_2}{2\pi}e^{i\vec{q}_t \cdot \vec{b}_2} F_i(b_2) S_i(\vec{b}_2-\vec{b}_1).
\label{eq:Sj}
\ee

\subsection{Generalisation to a two-channel eikonal}

Recall that in the two-channel eikonal, we take the Pomeron couplings to each diffractive eigen component $k$ to be\footnote{Following \cite{oldsoft}, we assume the same structure, that is the same shape and size for each component, apart, of course, from the cross section.}
\be
\beta_{P,k}(t)~=~(1\pm \gamma) \beta_P(t).
\ee
On the other hand, for the secondary Reggeons, $R$, which are believed to dominantly couple to valence quarks, we take
\be
\beta_{R,k}(t)~=~ \beta_R(t),
\ee
that is, the same $\beta_R$ for each component. Thus for the $PPP$, $RRP$ and $\pi\pi P$ couplings (with $j=P$) we make the replacement
\be
|I_d|^2~~\to~~ [(1+\gamma)|I_p|^2+(1-\gamma)|I_m|^2]/2
\label{eq:Idpi}
\ee
in (\ref{eq:3Rbscreen}), whereas for the $PPR$ and $RRR$ (with $j=R$) we let
\be
|I_d|^2~~\to~~ [|I_p|^2+|I_m|^2]/2.
\label{eq:IdR}
\ee
The subscripts $p,m$ are used to denote the $(1 \pm \gamma)$ eigen components respectively. In addition for $I_p$ we replace $S_i$, with $i=P$, in (\ref{eq:Sj}) by
\be
S_i~~\to~~S_{pP}~=~\frac{1}{2}\left[(1+\gamma)e^{-(1+\gamma)^2\Omega/2}+(1-\gamma)e^{-(1-\gamma^2)\Omega/2}\right]
\ee
and for $I_m$
\be
S_i~~\to~~S_{mP}~=~\frac{1}{2}\left[(1+\gamma)e^{-(1-\gamma^2)\Omega/2}+(1-\gamma)e^{-(1-\gamma)^2\Omega/2}\right].
\ee
These replacements of $S_i$ are for the $PPP$ and $PPR$ contributions. For the $RRP,~\pi\pi P$ and $RRR$ contributions, with $i=R$ or $\pi$, we have
\be
S_i~~\to~~S_{pR}~=~\frac{1}{2}\left[(e^{-(1+\gamma)^2\Omega/2}+e^{-(1-\gamma^2)\Omega/2}\right]
\label{eq:SpR}
\ee
\be
S_i~~\to~~S_{mR}~=~\frac{1}{2}\left[e^{-(1-\gamma^2)\Omega/2}+e^{-(1-\gamma)^2\Omega/2}\right].
\label{eq:SmR}
\ee

\subsection{The screening of the $\pi\pi P$ contribution}

We treat the $\pi\pi P$ contribution as a special case due to the presence of the $\pi$ spin-flip amplitude. First, neglecting screening, we have
\be
\frac{M^2 d\sigma}{dtdM^2}~=~\frac{G^2_{\pi\pi N}}{4\pi} \frac{-t}{(t-m^2_\pi)^2}\beta_P(0)g_{\pi\pi P}\left(\frac{s}{M^2}\right)^{2\alpha_\pi (t)-2}\left(\frac{M^2}{s_0}\right)^{\alpha_P(0)-1},
\label{eq:pi}\ee
where $G^2_{\pi\pi N}/4\pi=13.75$ \cite{pipiN}. Since the pion is almost on-mass-shell, we expect $g_{\pi\pi P} \simeq \frac{2}{3} \beta_P(0)$, according to the additive quark model. In other words 
\be
g_{\pi\pi P}/\beta_P(0)~\simeq~\sigma (\pi p)/\sigma(pp).
\ee
The factor $-t$ in (\ref{eq:pi}) is given by
\be
-t~=~q^2_t+q^2_\parallel~=~q^2_t+q^2_t\frac{1-x_L}{x_L}+\frac{m_p^2 (1-x_L)^2}{x_L},
\ee
where the $q_t^2=(\vec{q}_t \cdot \vec{\sigma})^2$ term is of proton spin-flip (sf) origin, and the remaining two terms are spin non-flip. Here, $\vec{\sigma}$ is the Pauli matrix associated with the spin of the incoming proton. The non-flip interaction, is screened in the usual way, see (\ref{eq:Idpi}), (\ref{eq:SpR}) and (\ref{eq:SmR}). However, the spin-flip amplitude is now a {\it vector} directed along $\vec{q}_t$. Thus the integrals in (\ref{eq:Fi}) and (\ref{eq:Sj}) should be written in vector form, with
\be
\frac{\beta_\pi^{\rm sf}(q_t)}{\beta_\pi^{\rm sf}(q_t=0)}~=~\frac{\vec{q}_t}{t-m_\pi^2},
\ee
and $(\beta_\pi^{\rm sf}(0))^2$ in (\ref{eq:A}) replaced by $G^2_{\pi\pi N}/4\pi$. This leads to
\be
{\vec F}_\pi^{\rm sf}(\vec{b}_2)~=~ \vec{b}_2  f_\pi^{\rm sf}(b_2)
\ee
and
\be
|I_d^{\rm sf}|^2~=~|I_x|^2+|I_y|^2
\ee
with $x$ directed along $\vec{q}_t$. After the azimuthal integration, it means that the Bessel function $J_0(qb)$, that arises in (\ref{eq:Fi}), should be replaced by $J_1(qb)$.

\section{Triple-Regge analysis of $pp \to pX$ and $\bar{p}p \to \bar{p}X$ data}
 
The main objective of this paper is to perform a triple-Regge analysis of the available $d^2\sigma/dtd\xi$ data for $pp \to pX$ and $\bar{p}p \to \bar{p}X$ (where $\xi=M^2_X/s$), allowing for screening (that is absorptive) effects. In this way we obtain a more reliable estimate of the {\it bare} triple-Pomeron coupling, $g_{PPP}$, which is a vital ingredient in the description of diffractive processes at high energy, see Ref.~\cite{GribMYF,GribM,KMRj}. As can be seen from the triple-Regge formalism of the previous section, it is necessary to work in impact parameter space in order to include screening corrections. However, it is computationally time-consuming to evaluate the required Fourier transforms inside a $\chi^2$ fit. To facilitate the fit to the data, we therefore first prepare grids of the integral in (\ref{eq:3Rbscreen}) for a range of fixed values of the slopes $b'_{iij}$ of the triple-Regge vertices for each $q_t$ and $M^2$ data point. A MINUIT fit to the data is then performed using Chebyshev polynomial interpolation in $b'_{iij}$.

In fact, we performed three triple-Regge fits to the $d^2\sigma/dtd\xi$ data with different assumptions for the form of the triple-Pomeron vertex. The first two assume either a strong or weak coupling triple-Pomeron vertex, that is they correspond to 
\be
g^S_{3P}(0)~~~~~ {\rm and} ~~~~~g^W_{3P}(0)q_t^2{\rm exp}(-b'q^2_t)
\ee
 respectively, see (\ref{eq:strong}) and (\ref{eq:weak}); recall that we use GeV units.  The third fit considers a combination of the two couplings with both $g^S_{3P}(0)$ and $g^W_{3P}(0)$ as free parameters.

We fit to CERN-ISR\footnote{We choose a subset of the ISR data which is sufficient to fully describe their $t$ and $\xi$ dependence.} \cite{jcma}, FNAL fixed-target \cite{rlc} and Tevatron \cite{fa} data for $pp \to pX$ and $\bar{p}p \to \bar{p}X$. The differential distributions for the FNAL fixed target and Tevatron experiments can be found in Ref.~\cite{GM}. The relative normalisations of the data sets are not very well known. In Ref.~\cite{GM} it was claimed that the normalisation uncertainties of the FNAL data are about 10$\%$. We fix the normalisation of the ISR data and introduce normalisation factors for the FNAL data with a 10$\%$ error. However, we find that the Tevatron data prefer to be normalised up by the strong coupling fit, and down for the weak coupling fit, by factors of about 25$\%$. We restrict this freedom and limit the normalisation factors to $\pm 15\%$ and $\pm 10\%$ for $\sqrt{s}=546$ and 1800 GeV data respectively, where the $+$ and $-$ signs refer to the strong and weak coupling fits respectively.
The remaining normalisations coming from the fits are $+10\%$ for the $\sqrt{s}=14,~20$ GeV data \cite{rlc}.

All three fits to the data prefer very small, or even negative, slopes, $b'_{iij}$, of the triple-Regge vertices. To avoid unphysical negative slopes, we impose the condition that $b'_{iij}>0$. In fact, the optimum $\chi^2$ for each fit has, within error bars, all $b'_{iij}=0$. Therefore, for the couplings quoted in Table 2, we have set all the slopes $b'_{iij}=0$.  The only exceptions to this are the slopes of the $PPP$ and $PPR$ vertices for the weak coupling fit, which are found to be positive, as shown in Table 2. We fixed the $\pi\pi P$ triple coupling at the additive quark model value, $g_{\pi\pi P}=\frac{2}{3} \beta_P (0)$. Actually the data prefer a smaller value, but with a large error. However the fixed value of $g_{\pi\pi P}$ only enlarges $\chi^2$ by less than 0.5. The couplings of the Reggeons to the proton are taken from the ``elastic'' analysis of Section 3.  The values of the remaining parameters, corresponding to the optimum triple-Regge fits, are given in Table 2.

\begin{table}[htb]
\begin{center}
\begin{tabular}{|c|c|c|c|}\hline
 & strong & weak & combination \\ \hline
$g^S_{3P}$  &   $0.44 ~\pm~0.05 $ &-& $0.44 ~\pm 0.10$  \\  
$g^W_{3P}$  &   - &$3.0 ~\pm~1.2 $& $0.2 ~\pm 0.5$  \\ 
$b^{\prime W}_{PPP}$  &   - &$1.15 ~\pm~0.3 $& -  \\ 
$g_{PPR}$  &   $0.75 ~\pm~0.10 $ &   $0.76 ~\pm~0.15 $&   $0.67 ~\pm~0.16 $  \\
$b^{\prime W}_{PPR}$  &   - &$1.4 ~\pm~1.7 $& -  \\ 
$g_{RRP}$  &   $1.1 ~\pm~0.3  $ &   $1.3 ~\pm~0.5  $&   $1.0 ~\pm~0.5  $ \\
$g_{RRR}$  &   $2.6 ~\pm~1.0  $ &   $2.9 ~\pm~1.4  $&   $2.8 ~\pm~1.5  $  \\ \hline
$\chi^2$/DoF & 0.83 & 1.40 & 0.83 \\ \hline
\end{tabular}
\end{center}
\caption{\sf The values of the {\it ``bare''} triple-Regge couplings $g_{iij}(0)$ of (\ref{eq:A}), and slopes $b'_{iij}$ of (\ref{eq:Fi}, \ref{eq:Fj}), obtained in the three optimum fits to the $d^2\sigma/dtd\xi$ data. Recall that all the slopes $b'_{iij}$ are set to zero, except for those of the $PPP$ and $PPR$ vertices in the weak coupling fit.}
\end{table}

\begin{figure} 
\begin{center}
\includegraphics[height=15cm]{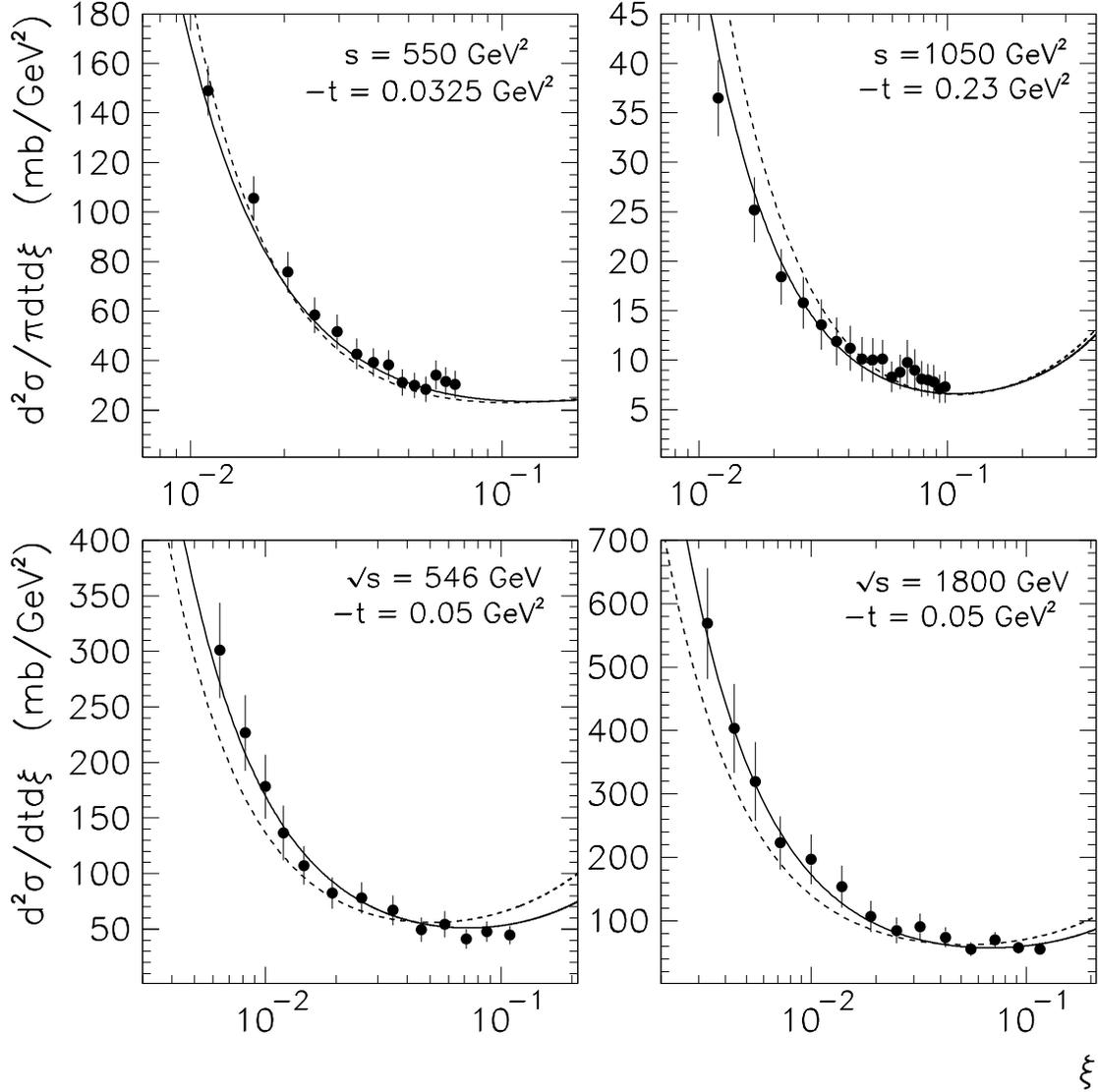}
\caption{\sf The description of a sample of the $d^2\sigma/dtd\xi$ cross section data that are fitted  using the strong (continuous curves) and weak (dashed curves) triple-Pomeron coupling ansatzes. ($\xi\simeq M^2/s$). Here, the curves corresponding to the (strong, weak) coupling fits of the FNAL data have been normalised (down, up) by 15$\%$ at $\sqrt{s}=$ 546 GeV and by 10$\%$ at $\sqrt{s}=$ 1800 GeV, to allow for the normalisations found for these data in the respective fits.}
\label{fig:weak}
\end{center}
\end{figure}

We see from the $\chi^2$ per degree of freedom (DoF) that the data clearly prefer the {\it strong}, rather than the {\it weak}, triple-Pomeron coupling ansatz. This is also clear from the ``weak+strong combination'' fit, which is dominated by the strong coupling component, and is little changed from the pure strong coupling fit\footnote{The normalisations of the data that are found in the ``combination'' fit are the same as those of the strong coupling fit.}. The preference for the strong triple-Pomeron coupling is also evident from Fig.~\ref{fig:weak}. This compares the ``strong'' and ``weak'' descriptions of a sample of the $d^2\sigma/dtd\xi$ data that are  fitted, including especially the high energy FNAL data at $\sqrt{s}=$ 546 and 1800 GeV. The FNAL data are most relevant since the triple-Pomeron contribution dominates for $\xi \lapproxeq 0.03$. We see that the weak coupling description of the $\xi$ shape of these data is disfavoured. 

\begin{figure} 
\begin{center}
\includegraphics[height=17cm]{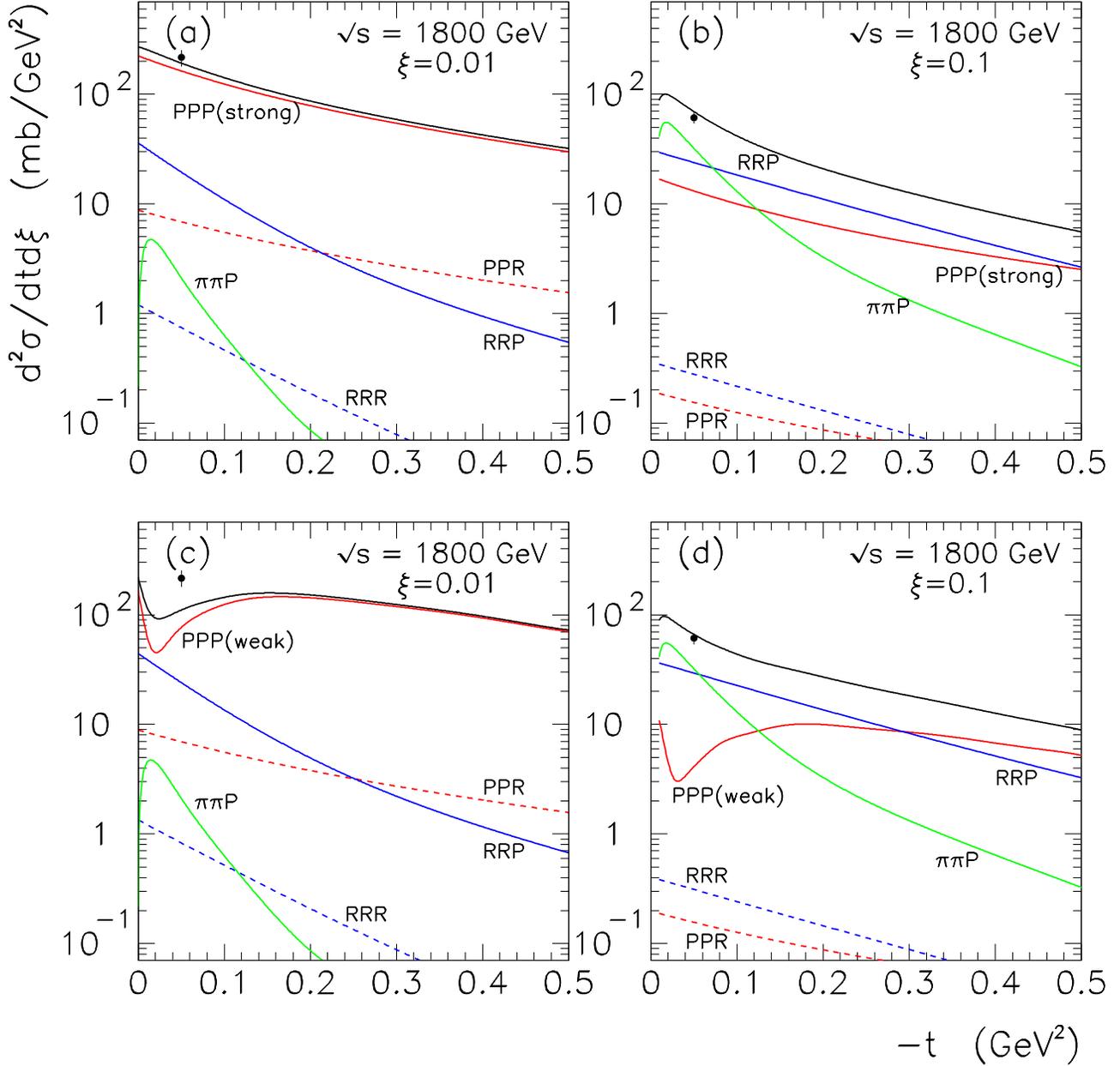}
\caption{\sf The four plots show the $t$-dependence of the $d^2\sigma/dtd\xi$ at $\xi=0.01,~0.1$ and $\sqrt{s}=1800$ GeV obtained in the strong and weak triple-Pomeron fits respectively.}
\label{fig:tdep}
\end{center}
\end{figure}

\begin{figure} 
\begin{center}
\includegraphics[height=17cm]{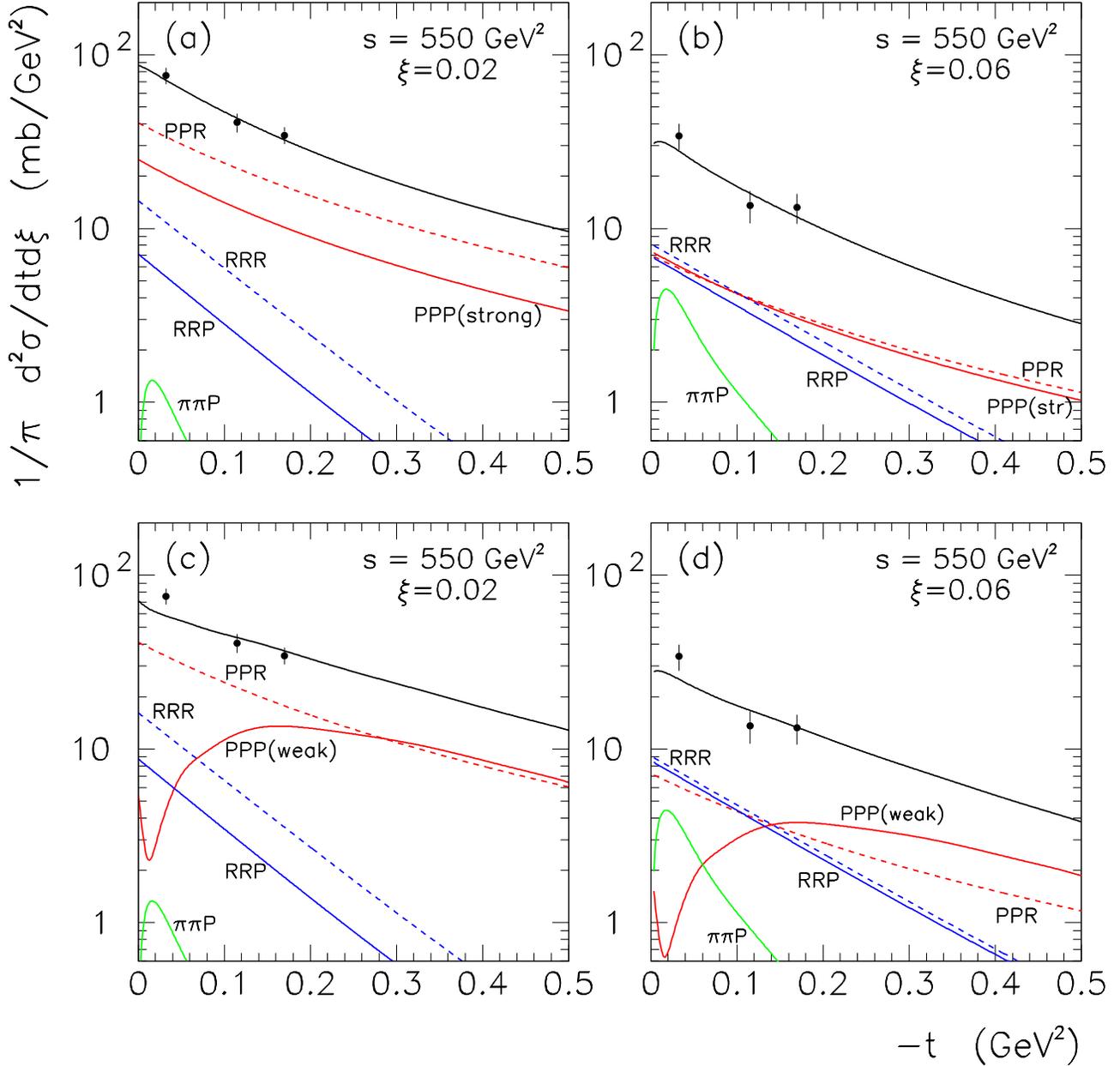}
\caption{\sf The four plots show the $t$-dependence of the $d^2\sigma/dtd\xi$ at $\xi=0.02,~0.06$ and $s=550 ~{\rm GeV}^2$ obtained in the strong and weak triple-Pomeron fits respectively, together with the data available at these kinematic values.}
\label{fig:tdep2}
\end{center}
\end{figure}

\begin{figure} 
\begin{center}
\includegraphics[height=8cm]{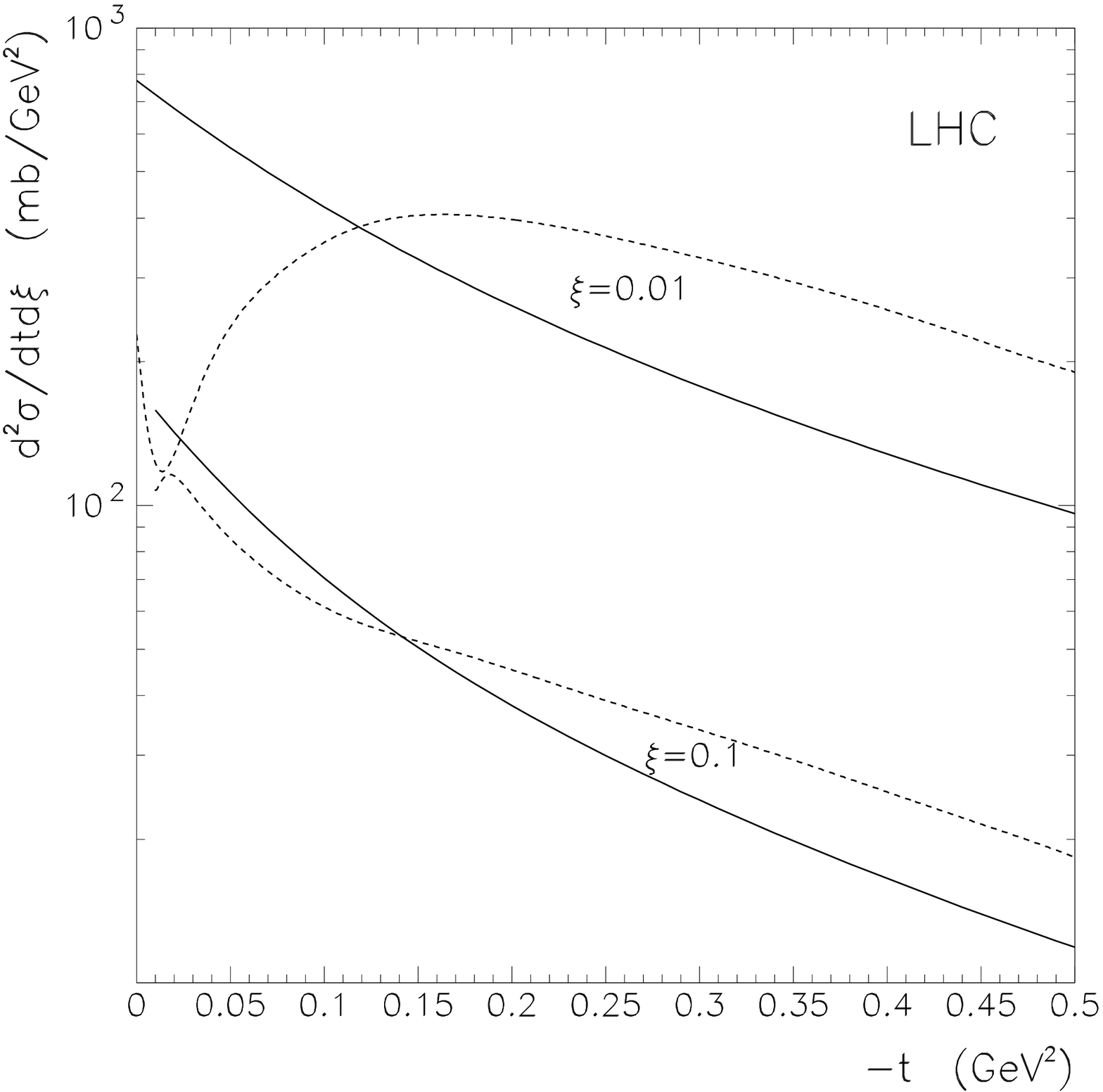}
\caption{\sf The continuous curves are the predictions for the $t$-dependence of the $d^2\sigma/dtd\xi$ at $\xi=0.01,~0.1$ and $\sqrt{s}=14$ TeV obtained in the strong triple-Pomeron fit. The disfavoured weak coupling predictions are shown by dashed curves.}
\label{fig:tdepLHC}
\end{center}
\end{figure}

The $t$-dependence of the single diffractive cross section, $d^2\sigma/dtd\xi$, is shown, for example,
in Fig.~\ref{fig:tdep} for $\xi=0.01,~0.1$ at $\sqrt{s}=1800$ GeV, and in Fig.~\ref{fig:tdep2} for $\xi=0.02,~0.06$ at the much lower energy corresponding to $s=550~ {\rm GeV}^2$. We also show the individual triple-Regge contributions in these plots, together with the data available at these kinematic values.  Note that, after accounting for the absorptive effects, the
triple-Pomeron contribution does not vanish in the forward direction even in the {\it weak} coupling fit. Therefore to choose between the
{\it weak} and {\it strong} coupling scenarios we have had to perform a full
fit of the data. It is not enough just
to study the low $q_t$ behaviour of the cross section. The first indication of the weak coupling characteristics is seen in Fig.~\ref{fig:tdep} for $\xi \sim 0.01$ and $|t| <0.1 ~{\rm GeV}^2$ at the Tevatron energy. In this domain the data favour the strong coupling scenario.  For larger values of $\xi$ the dip produced by the $PPP$ term in the weak coupling case is filled in by the peak due to the $\pi\pi P$ term. At the lower energy, corresponding to 550 ${\rm GeV}^2$, we see, from Fig.~\ref{fig:tdep2}, that it is difficult to distinguish between the two scenarios. However, at the higher energy of the LHC the weak coupling scenario will reveal itself by a well pronounced dip at $t \simeq -0.02~ {\rm GeV}^2$ after which $d^2\sigma/dtd\xi$ increases with $|t|$ up to $|t| \simeq 0.15~ {\rm GeV}^2$, see Fig.~\ref{fig:tdepLHC}.
 It is clear from Fig.~\ref{fig:tdepLHC} that single diffractive dissociation data obtained at the LHC for $\xi \sim 0.01$ should be able to readily distinguish between the strong and weak triple-Pomeron coupling scenarios.

\begin{figure} 
\begin{center}
\includegraphics[height=14cm]{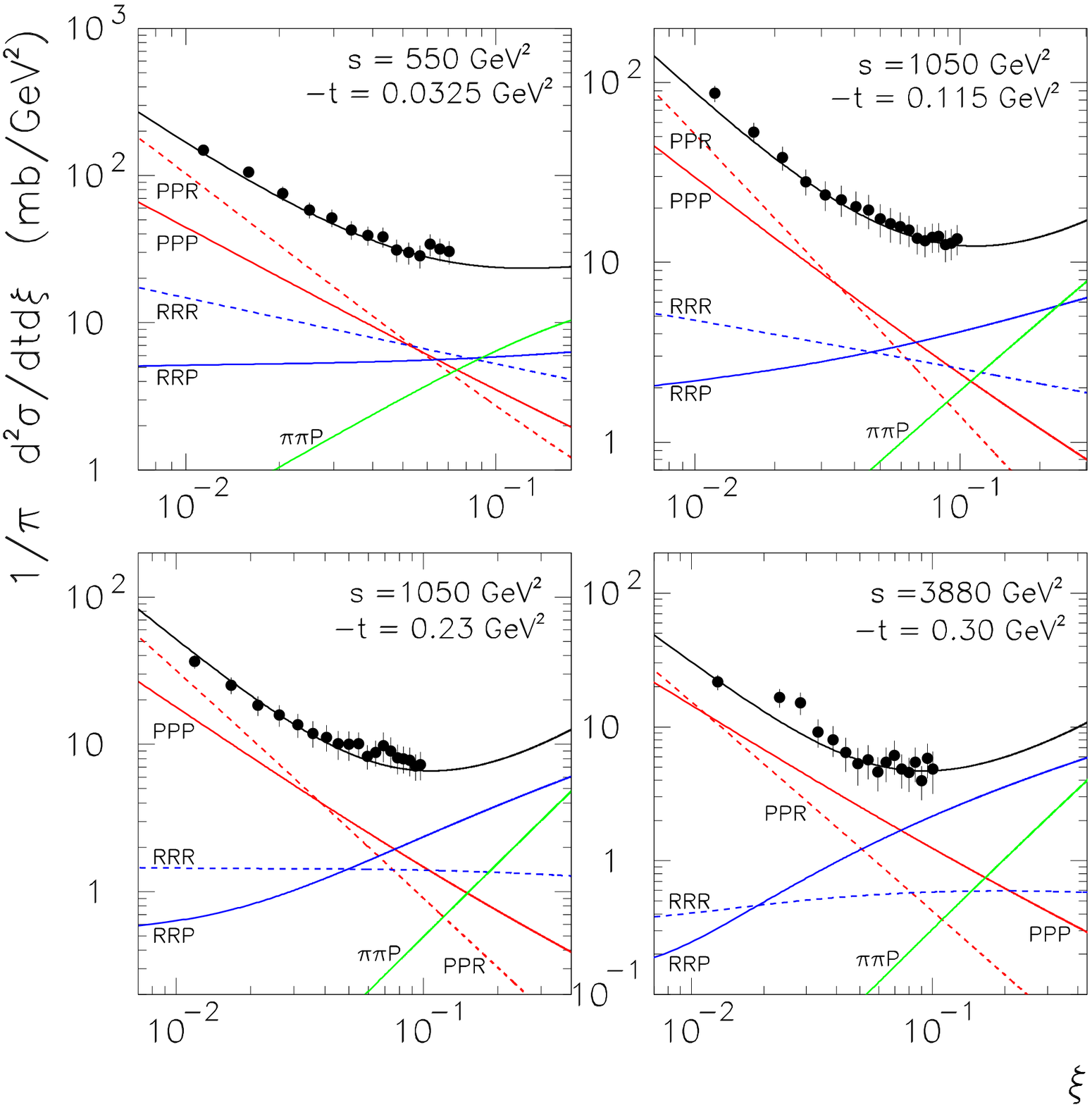}
\caption{\sf The description of the CERN-ISR $pp \to pX$ cross section, $d^2\sigma/dtd\xi$, data \cite{jcma} obtained in the strong triple-Pomeron coupling fit. ($\xi\simeq M^2/s$). The individual triple-Regge contributions are also shown.}
\label{fig:ISR}
\end{center}
\end{figure}

\begin{figure} 
\begin{center}
\includegraphics[height=14cm]{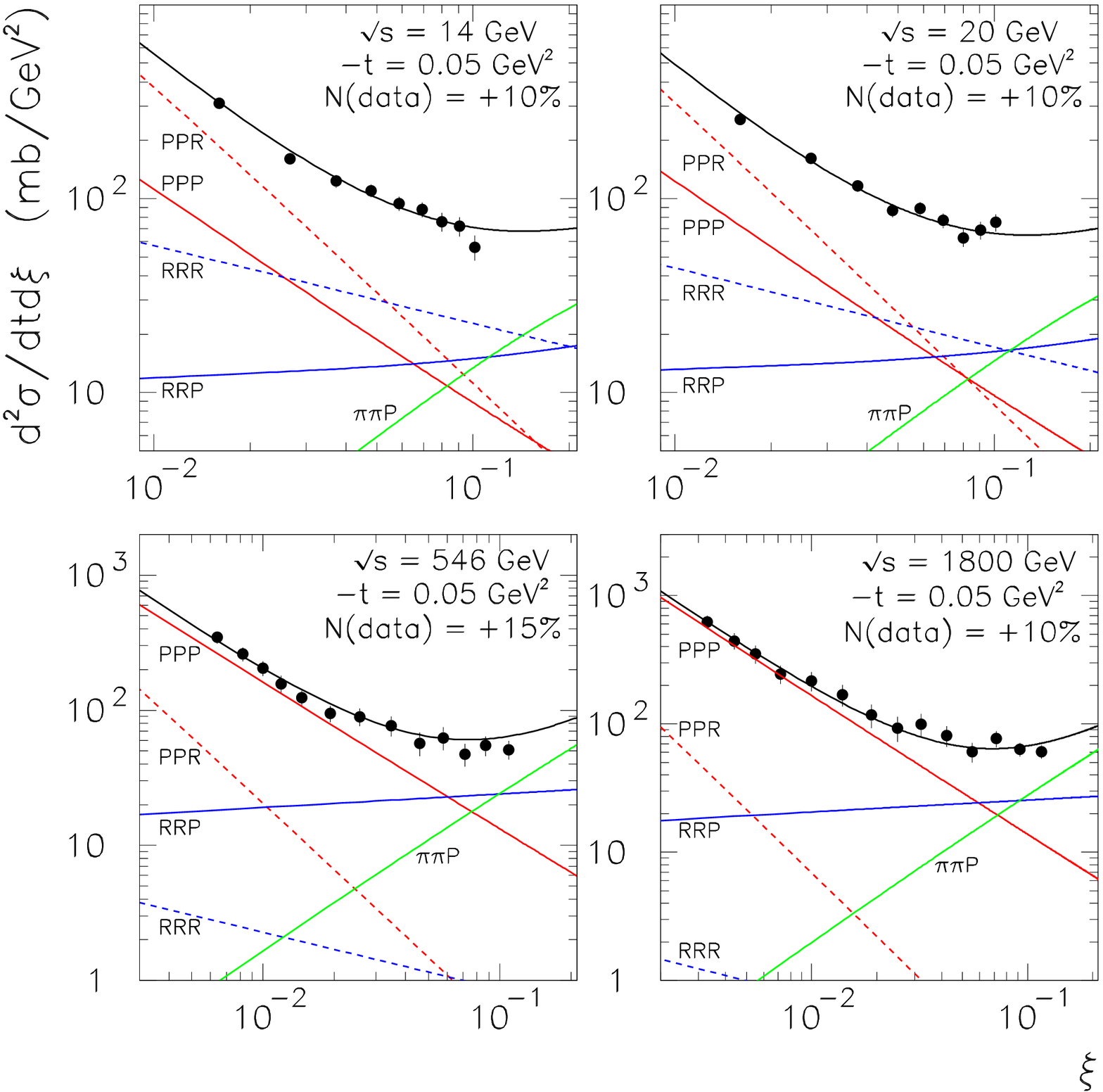}
\caption{\sf The description of the $d^2\sigma/dtd\xi$, measured in fixed-target and collider experiments at FNAL \cite{rlc,fa,GM}, obtained in the strong triple-Pomeron coupling fit. The individual triple-Regge contributions are also shown.}
\label{fig:Tev}
\end{center}
\end{figure}

In the remainder of this Section, we discuss only the favoured strong triple-Pomeron coupling fit, for which the triple-Regge couplings are the first set listed in Table 2.  
The values of these couplings are rather stable. They never go outside the quoted errors when we change the renormalisation parameters of the data, or allow $g_{\pi\pi P}$ to be a free parameter. The quality of the fit is excellent, with a minimum $\chi^2/{\rm DoF}=167/(210-8)=0.83$. The 8 parameters are the 4 couplings, $g_{iij}(0)$, and the 4 data renormalisations. The quality of the description can been seen from the plots of samples of the data that are fitted, Fig.~\ref{fig:ISR} and Fig.~\ref{fig:Tev}, which also show the various triple-Regge contributions to the cross section.

In comparison with the old triple-Regge analysis of Ref.~\cite{FF}, we now
obtain a more than twice larger relative contribution of the $PPR$
term. This is mainly due to the inclusion in our analysis
of the higher energy Tevatron data. The inclusive cross sections
measured in the interval of $\xi\sim 0.01 - 0.03$ at the
Tevatron energies turn out to be about twice smaller than that
measured at the low CERN-ISR energies.

\begin{figure} 
\begin{center}
\includegraphics[height=4cm]{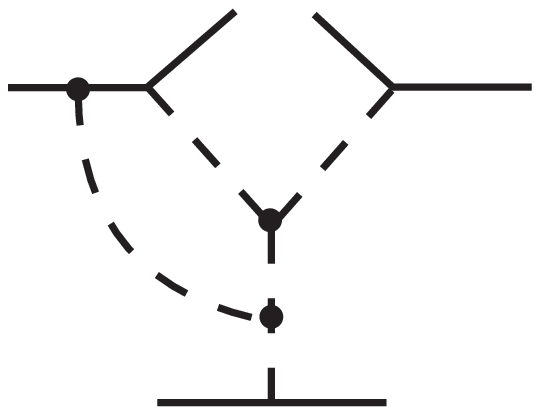}
\caption{\sf An example of an ``enhanced'' contribution to high-mass diffraction.}
\label{fig:en}
\end{center}
\end{figure}

The ``strong coupling'' parameters of Table 2 are much closer to the {\it bare} triple-Reggeon couplings than those coming from the old fits \cite{KKPI,FF,Kaid} which did not allow for the screening corrections. In particular the value\footnote{Note that this result is in reasonable agreement with the parametrisation of the renormalised triple-Pomeron amplitude of Ref.~\cite{GM} which leads to $g_{3P} ~\simeq ~0.15~\beta_P$}.
\be
g_{3P}~ \equiv ~g_{PPP}~\simeq~0.2~\beta_P
\ee
is consistent with the reasonable extrapolation of the perturbative BFKL Pomeron vertex to the low scale region \cite{brv}. However, these are still not the true bare vertices. For these sizeable values of $g_{iij}$ the effect of ``enhanced'' screening, shown in Fig.~\ref{fig:en}, is not 
negligible. Moreover, it is not enough to just take one iteration, that is, to repeat the ``elastic'' analysis, but now including the effects of the triple-Regge high-mass absorption, obtain new Regge-proton parameters and then repeat the triple-Regge analysis\footnote{Note that an eikonal-type model for absorption will still violate $s$-channel unitarity \cite{ms}. At large $b$, on the periphery, where the opacity $\Omega (b)$ is small, the effect of eikonal screening is not effective. For these (large $l$) partial waves the contribution coming from diffractive dissociation becomes larger than the total inelastic contribution. To satisfy unitarity more complicated (enhanced) multi-Pomeron diagrams must be included.}. Rather, it is necessary to sum up the series of multi-Reggeon diagrams. We have considered such a model \cite{newnewsoft}, that is an extension of an earlier model \cite{newsoft} but now including non-zero $\alpha'_P$ as a free parameter, as well as more $t$-channel exchanges. The tuning of this model shows that it is possible to obtain a good description of the data provided that the triple-Pomeron coupling is a bit larger
\be
g_{3P} ~\simeq ~0.25~\beta_P.
\ee

\section{Inelastic $J/\psi$ photoproduction}

In Ref. \cite{KMRj} it was pointed out that the observation of the process $\gamma p \to J/\psi+Y$ at large values of $M_Y$ offers, in principle, an opportunity to determine the triple-Pomeron coupling where the screening corrections are smaller than in the pure hadronic reactions, see also \cite{GLM07}. Unfortunately, the $M_Y^2$ distribution has not been measured yet.  However there exists a comparison of the HERA data for the ``elastic'' photoproduction process, $\gamma p \to J/\psi+p$ with the proton dissociation data. The ratio, at the photon-proton centre-of-mass energy $W=200$ GeV and $t=0$, is \cite{549}-\cite{Akt}.
\be
r~\equiv~\frac{d\sigma(\gamma p \to J/\psi+Y)/dt}{d\sigma(\gamma p \to J/\psi+p)/dt}~\simeq~0.2,
\label{eq:r}
\ee
where the ``inelastic'' cross section has been integrated over the mass region $M_Y<30$ GeV. Since only the Pomeron couples to charm quarks, the ratio $r$ is described by the $PPP$ and $PPR$ contributions. The cross section for $J/\psi$ absorption is rather small. So we may neglect the screening factor, and obtain
\be
r~=~\frac{1}{\pi \beta_P(0)}\int\frac{dM^2}{M^2}\left(\frac{s_0}{M^2}\right)^{2{\hat \alpha}_P-2}\left[g_{PPP}\left(\frac{M^2}{s_0}\right)^{\alpha_P(0)-1}+~g_{PPR}~\frac{\beta_{f_2}(0)}{\beta_P(0)}\left(\frac{M^2}{s_0}\right)^{\alpha_R(0)-1}\right].
\label{eq:trf}
\ee

\begin{figure}
\begin{center}
\includegraphics[height=5cm]{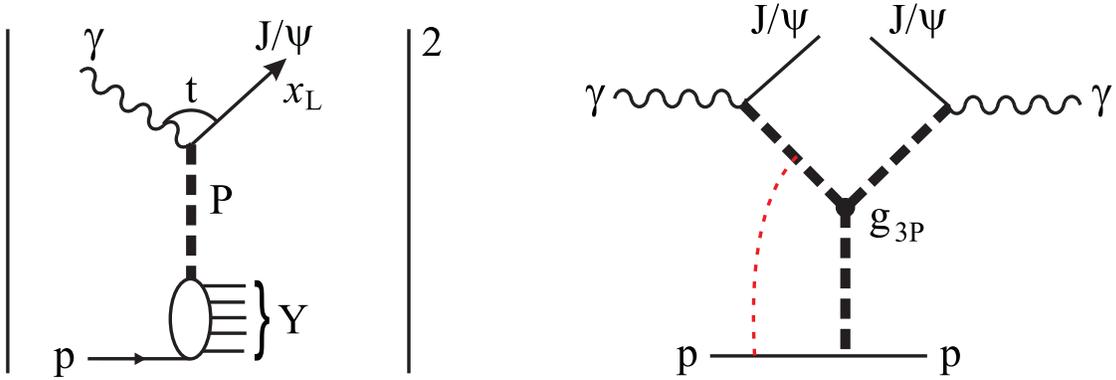}
\caption{\sf The process of proton dissociation in diffractive $J/\psi$ photoproduction, $\gamma+p \to J/\psi+Y$, which is described by a diagram with a triple-Pomeron vertex in which the rescattering effects are small. The dotted line would mean the diagram became an enhanced diagram. This contribution is small. }
\label{fig:2}
\end{center}
\end{figure}
Note that the Pomeron trajectories $\alpha_P(0)$ and ${\hat \alpha}_P(t)$ in (\ref{eq:trf}), that is in the triple-Pomeron diagram in Fig.~\ref{fig:2}, are not the same. The lower Pomeron $\alpha_P(0)$ in Fig.~\ref{fig:2} is the usual `soft' Pomeron; whereas the upper ones, with ${\hat \alpha}_P(t)$,
include DGLAP evolution from a low initial scale $\mu=\mu_0$ up to a rather large scale $\mu\sim M_{J/\psi}$ at the $J/\psi$
production vertex. The summation of the double logarithms $(\alpha_s\ln(1/x)\ln(\mu^2/\mu^2_0))^n$ leads to a steeper $x$-dependence and hence to a larger effective intercept
for the trajectory ${\hat \alpha}_P(t)$ of the upper `hard' Pomeron. Thus to evaluate the ratio $r$ of (\ref{eq:trf}) we take ${\hat \alpha}_P=1.18$, which corresponds to the $W$ dependence observed in the HERA data \cite{zeus695}-\cite{Akt}. Taking the other parameters from Table 1 and the strong coupling fit of Table 2, we obtain for the two terms in (\ref{eq:trf})
\be
r~\equiv~r_{PPP}+r_{PPR}~=~0.12+0.06.
\ee
This is consistent with the HERA data, within the uncertainties. 

We also use the weak triple-Pomeron coupling fit of Table 2 to evaluate the ratio $r$ at $t=-0.2~{\rm GeV}^2$. This is the lowest $-t$ value for which $r$ is measured. The experimental value is\footnote{The measured ratio at $t=-0.2~{\rm GeV}^2$ is larger than that of (\ref{eq:r}) at $t=0$, since the cross section of the elastic process, $\gamma p \to J/\psi + p$, has a larger $t$-slope than that for $\gamma p \to J/\psi + Y$.} \cite{549}
\be
r~=~0.4 \pm 0.1,~~~~~~~{\rm for}~~t=-0.2~{\rm GeV}^2.
\ee
When we evaluate the weak $PPP$ coupling and $PPR$ contributions to $r$, we find
\be
r~\equiv~r_{PPP}+r_{PPR}~=~0.14+0.07,~~~~~~~{\rm for}~~t=-0.2~{\rm GeV}^2,
\ee
a value about a factor two smaller than that observed. So again the strong triple-Pomeron coupling is favoured over the weak.

Originally, the strong triple-Pomeron coupling was supported by an analysis \cite{KMRj} of the HERA $J/\psi$ photoproduction data. The problem is that, at present, the $M_Y^2$ dependence of the cross section has not been measured; only the value integrated over an $M_Y^2$ interval is given. The large value of the triple-Pomeron coupling that was extracted in \cite{KMRj} assumed that the whole cross section originates from Pomeron exchange, and neglected the secondary Reggeon contribution. Here we have included the secondary Reggeon contribution. However, if the $M_Y^2$ distribution were to be observed, it would then be possible to separate these contributions using just the $J/\psi$ photoproduction data, and to give an independent check of the results of our triple-Regge analysis.
Clearly a dedicated measurement of the process $\gamma p \to J/\psi+Y$, including especially the $M_Y$ and $t$ dependence, would be extremely informative.

\section{Conclusions}

We have performed a triple-Regge analysis of the available
$pp\to p+X$ and $\bar pp\to \bar p +X$ data, in which we account for
absorptive effects. Thus, the couplings of the triple-Reggeon vertices extracted in this
analysis will be much closer to the couplings of the original {\it bare} vertices than
those obtained in the old analyses, where the screening corrections
were implicitly included in the values of {\it effective} couplings.

The triple-Pomeron vertex turns out to be rather large,
 $g_{3P}\simeq 0.2\beta_P$. This indicates that, in a more precise
description of ``soft'' interactions, we must include more complicated 'enhanced' diagrams and diagrams with the multi-Pomeron vertices (that is with a larger number of Pomerons
in a vertex).

Due to the inclusion of the high-energy Tevatron data in the analysis,
we now obtain a twice larger relative contribution of the $PPR$ term
in comparison with the results of the old analyses.

We considered two possible parametrisations
of the small $q_t$ behaviour of the triple-Pomeron vertex:  
\be
g_{3P}={\rm constant} ~~~~~~~~ {\rm and} ~~~~~~~~g_{3P}=g^W_{3P}(0)q^2_t{\rm exp}(-b'q^2_t), 
\ee
in GeV units. We found that the data prefer the first, the
so-called {\it strong} Pomeron coupling scenario.  In the second, the so-called {\it
weak} coupling, case, we obtained a much poorer description of the high-energy Tevatron (CDF)
data at small $\xi$ and low $-t=0.05$ GeV$^2$, where the triple-Pomeron
contribution dominates.  It will be important
to measure the inclusive cross section for single diffraction, $d^2\sigma/dtd\xi$, at the LHC to
confirm this conclusion. This could be done when the forward detectors are operating at the LHC, even at moderate integrated luminosity \cite{early}. Another possibility
is to study in more detail $J/\psi$ diffractive production with
dissociation of the target proton. That is, to measure the $M^2_Y$ and $t$
dependence of the $\gamma p\to J/\psi + Y$ reaction. The 
HERA experiments have so far only presented the cross section integrated over $M^2_Y$ up
 to $M_Y=30$ GeV.  However, already these data support the {\it strong}
coupling solution.

Finally, we note that our analysis of the data prefers zero slopes, corresponding to
small size of the bare triple-Reggeon vertices. Together with the larger
values of the $PPP$ and $PPR$ vertices, this gives hope that there is a smooth
matching to the perturbative QCD treatment of the Pomeron.

\section*{Acknowledgements}

We thank Aliosha Kaidalov, Alessia Bruni and Risto Orava for useful discussions.
MGR and EGSL thank the IPPP at the University of Durham for hospitality.
The work was supported by INTAS grant 05-103-7515, by grant RFBR
07-02-00023, by the Russian State grant RSGSS-3628.2008.2 and by the CNPq (Brazil) under contract 210242/2006-0.

\end{document}